\documentclass[11pt]{article}

\newfont{\goth}{eufm10 scaled\magstep1}

\let\ssection=\section
\renewcommand{\section}{\setcounter{equation}{0}\ssection}

\def\parag{\hfil\break} 
\def\kikezd{\parag\underbar}
\def\p{{\partial}}

\def\vP{{\vec P}}

\def\vE{{\vec E}}
\def\vX{{\vec X}}
\def\cP{{\cal P}}
\def\cA{{\cal A}}
\def\cE{{\cal E}}
\def\cJ{{\cal J}}

\def\cK{{\cal K}}
\def\cC{{\cal C}}

\newcommand{\gothg}{\mbox{\goth g}}
\newcommand\G{\widetilde{\gothg}}

\newcommand\half{{\scriptstyle{\frac{1}{2}}}}

\begin{document}

\setlength{\baselineskip}{16pt}

\title{Enlarged Galilean symmetry of anyons and the Hall effect}

\author{
P.~A.~Horv\'athy\footnote{e-mail: horvathy@univ-tours.fr}
\\
Laboratoire de Math\'ematiques et de Physique Th\'eorique\\
Universit\'e de Tours\\
Parc de Grandmont\\
F-37 200 TOURS (France)
\\
L. Martina\footnote{e-mail: Luigi.Martina@le.infn.it}
\\
Dipartimento di Fisica dell'Universit\`a
\\
and\\
Sezione INFN di Lecce. Via Arnesano, CP. 193\\
I-73 100 LECCE (Italy)
\\ and\\
P.~C.~Stichel\footnote{e-mail: peter@Physik.Uni-Bielefeld.DE}
\\
An der Krebskuhle 21\\
D-33 619 BIELEFELD (Germany)
}

\date{\today}

\maketitle

\begin{abstract}
Enlarged planar Galilean symmetry, built of both space-time and field 
variables and also incorporating the  ``exotic'' central extension is 
introduced. It is used to describe
 non-relativistic anyons coupled to an electromagnetic field. Our 
theory exhibits an anomalous  velocity relation of the type used to 
explain the  Anomalous Hall Effect. The Hall motions, characterized 
by a Casimir of the enlarged algebra,  become mandatory for some 
critical value(s) of the magnetic field. The extension of our scheme 
yields the semiclassical effective model of the Bloch electron.
\end{abstract}


\section{Introduction}

The planar Galilei group admits a
two-fold ``exotic'' central extension, labeled with
$m$ (the mass) and a second, ``exotic'' parameter $\kappa$ 
\cite{exotic}.
Models which provide a physical realization
of this ``exotic'' symmetry have been presented in \cite{LSZ1,DH}.
  Below, we focus our attention on the theory of \cite{DH}, since
that of \cite{LSZ1} is in fact an extended version of the latter.
Minimal (symplectic) coupling of the particle to an external 
electromagnetic field yields the first-order phase space Lagrangian
\begin{equation}
     L=
     P_{i}\cdot\dot{X}_{i}-\frac{\vP^2}{2m}
     +e(A_{i}\dot{X}_{i}+A_{0})
     +\frac{\theta}{2}\epsilon_{ij}P_{i}\dot{P}_{j},
     \label{totlag}
\end{equation}
where $\theta=-\kappa/m^2$ is the non-commutative parameter.
The Euler-Lagrange equations are
\begin{eqnarray}
     m^* 
\dot{X}_{i}&=&P_{i}-em\theta\epsilon_{ij}E_{j},\label{vitesse}
     \\[4pt]
     \dot{P}_{i}&=&e 
B\epsilon_{ij}\dot{X}_{j}+e\,E_{i},\label{Lorentz}
\end{eqnarray}
where $E_{i}$ and $B$ are the electric and magnetic field,
respectively, and
$
m^*=m(1-e\theta B)
$
is an effective mass. These equations can also be obtained
in a Hamiltonian framework, using the usual Hamiltonian
$
H=\vP{}^2/2m-eA_{0}
$
and the modified Poisson-brackets
\begin{equation}
     \begin{array}{ccc}
	\{X_{i},X_{j}\}=\displaystyle\frac{m}{m^*}\,\theta\epsilon_{ij},
	&\{X_{i},P_{j}\}=\displaystyle\frac{m}{m^*}\,\delta_{ij},
	&\{P_{i},P_{j}\}=\displaystyle\frac{m}{m^*}\,eB\epsilon_{ij}.
     \end{array}
     \label{Bcommrel}
\end{equation}
The most dramatic prediction of the model is that
when $m^*=0$ i. e., when the magnetic field takes the critical value
\begin{equation}
     B=B'_{crit}=\frac{1}{e\theta},
     \label{critB}
\end{equation}
the system becomes singular, and the only allowed motions
  follow the Hall law \cite{DH}. 
  The Poisson-brackets (\ref{Bcommrel}) are changed for new ones in a
reduced phase-space (see \cite{DH}.

   Requiring $B=B'_{crit}$ amounts to a
restriction to the lowest Landau level, and quantization allowed
us to recover the ``Laughlin'' wave functions \cite{DH}.

In this Letter we generalize this model,
and indicate its relation to models used in
solid state physics \cite{AHE, Niu}.

\section{Enlarged Galilean symmetry}\label{enlarged}

Adapting the idea of \cite{NeOl} to planar physics, we
consider a homogeneous electric field
$E_{i}(t)$ and a constant magnetic field $B$, and view $E_{i}$
and its canonical conjugate momentum, $\pi_{i}$, as additional
variables on an enlarged phase space.
This latter is endowed with  the enlarged Lagrangian
$
L^{enl}=L+\pi_{i}\dot{E}_{i}.
$
The $\pi_{i}$ are Lagrange multipliers and one of the
equations of motion is
$
\dot{E}_{i}=0,
$
i.e., the electric field should actually be constant.

The Galilean symmetry
of combined particle+homogeneous field system is readily established:
the enlarged Lagrangian  is (quasi-)invariant w. r. t. enlarged
  space translations, rotations and boosts, implemented as
\begin{equation}
     \begin{array}{cccc}
    \delta X_{i}=a_{i},
    \qquad
    &\delta P_{i}=0,
    \qquad
    &\delta E_{i}=0,
    \qquad
    &\delta\pi_{i}=ea_{i}t,\hfill
\\[6pt]
    \delta X_{i}=-\varphi\, \epsilon_{ij}X_{j},
    \qquad
    &\delta P_{i}=-\varphi\, \epsilon_{ij}P_{j}
    \qquad
    &\delta E_{i}=-\varphi\, \epsilon_{ij}E_{j},
    \qquad
    &\delta\pi_{i}=-\varphi\, \epsilon_{ij}\pi_{j},\hfill
\\[6pt]
    \delta X_{i}=b_{i}t,
    \qquad
    &\delta P_{i}=mb_{i},
    \qquad
    &\delta E_{i}=-B\epsilon_{ij}b_{j},
    \qquad
    &\delta\pi_{i}=eb_{i}\frac{t^2}{2}.\hfill
\end{array}
\label{engal}
\end{equation}

Now we consider the enlarged Hamiltonian structure.
Then the equations of motion
(\ref{vitesse})-(\ref{Lorentz}), augmented with
$
\dot{\pi}_{i}=eX_{i}
$
are Hamiltonian, $\dot{Y}=\{Y,H\}$, with the usual
  Hamiltonian
$
H=\vP^2/2m-eE_{i}X_{i},
$
and the fundamental Poisson brackets
(\ref{Bcommrel}),
supplemented with $\{E_{i},\pi_{j}\}=\delta_{ij}$.
  Then conserved quantities are readily constructed.
Integration of the equation of motion
(\ref{Lorentz}) shows that
\begin{equation}
     {\cal P}_{i}=P_{i}-eB\epsilon_{ij}X_{j}-eE_{i}t
    \label{enlargedmom}
\end{equation}
is a constant of the motion. Using the commutation relations
\begin{equation}
     \{X_{i},\cP_{j}\}=\delta_{ij},
     \qquad
     \{P_{i},\cP_{j}\}=0,
     \qquad
     \{\cP_{i},E_{j}\}=0
     \qquad
     \{\pi_{i},\cP_{j}\}=et\,\delta_{ij},
\end{equation}
  we find furthermore that (\ref{enlargedmom})
  generates enlarged translations, (\ref{engal}).
Similarly,
\begin{eqnarray}
     \cJ=\epsilon_{ij}X_{i}P_{j}+
     \frac{\theta}{2}{\vP}^2+
     \frac{eB}{2}{\vX}^2+\epsilon_{ij}E_{i}\pi_{j}+s_{0}
     \label{enlargedangmom}
     \\[6pt]
     \cK_{i}=mX_{i}-\left(\cP_{i}+e\frac{E_{i}t}{2}\right)t
     +m\theta\epsilon_{ij}P_{j}-B\epsilon_{ij}\pi_{j}
     \label{enlargedboostCC}
\end{eqnarray}
are conserved and generate enlarged rotations and  boosts,
respectively.
Note that anyonic spin, represented by the real number
$s_{0}$, has also been included.
%
%
  The  generators satisfy the enlarged Poisson relations
\begin{equation}
     \begin{array}{ccc}
	\{\cP_{i},H\}=eE_{i},\hfill
	&\{\cK_{i},H\}=\cP_{i},\hfill
	&\{\cJ,H\}=0,\hfill
	\\[10pt]
	\{\cP_{i},\cJ\}=-\epsilon_{ij}\cP_{j},\hfill
	&\{\cK_{i},\cJ\}=-\epsilon_{ij}\cK_{j},\hfill
	&\{\cP_{i},\cK_{j}\}=-m\delta_{ij},\hfill
         \\[10pt]
	\{\cP_{i},\cP_{j}\}=-eB\,\epsilon_{ij},\quad\hfill
	&\{\cK_{i},\cK_{j}\}=-m^2\theta\,\epsilon_{ij}.\quad\hfill&
	\end{array}
	\label{enlargedGal}
\end{equation}
A  closed algebra is obtained, therefore,  if
the electromagnetic fields $E_{i}$ and $B$ are conside\-red
as additional  elements of an enlarged Galilei
algebra $\G$, cf. \cite{NeOl}.
  The (constant) magnetic field, $B$, belongs, together with
  $m$ and $\kappa=-\theta m^2$, to the  center of $\G$.
The additional nonzero brackets are
\begin{equation}
	\{E_{i},\cJ\}=-\epsilon_{ij}E_{j},
	\qquad
	\{E_{i},\cK_{j}\}=B\,\epsilon_{ij}.
	\label{more}
\end{equation}

Our enlarged Galilei group has two independent Casimirs, namely
\begin{eqnarray}
     \cC\hfill&=&e\theta\left(BH-\epsilon_{ij}\cP_{i}E_{j}
     +\displaystyle\frac{m}{2B}\vE^2\right),\hfill
     \label{casimir2}
     \\[6pt]
     \cC'\hfill&=&\displaystyle\frac{{\vec{\cP}}^2}{2m}-H-\frac{e}{m}
     \Big(\cK_{i}E_{i}+\cJ B\Big)-
     \displaystyle\frac{me\theta}{2B}\vE^2,\hfill
     \label{casimir1}
\end{eqnarray}
In the representation of the enlarged Galilei algebra given in terms of
phase-space variables,
\begin{eqnarray}
     \cC\hfill=\displaystyle\frac{e\theta B}{2m}
     \Big(P_{i}-\displaystyle\frac{m}{B}\epsilon_{ij}E_{j}\Big)^2
     \qquad\hbox{and}\qquad
     \cC'=-\cC-\frac{es_{0}B}{m}.
     \label{Casimir}
\end{eqnarray}

$\cC'$ generalizes
one of the two Casimirs of the  planar Galilei group, namely
the internal energy \cite{galcasi}. $\cC+\cC'$ is in turn proportional
to the second Casimir identified as the spin;
these two quantities are linked, just like
for the model of \cite{LSZ1}. The relation of the
enlarged and ordinary Galilean algebras can be clarified by
a subtle group contraction.
Though constants of the motion,
our Casimirs are not fixed constant~:
the representation of our enlarged Galilei group is, in general 
reducible.
If the fields become non-dynamical,
those transformations which are consistent
with the constant fields remain symmetries.
For example, (\ref{enlargedmom}) becomes the
familiar magnetic translation.

\section{Algebraic construction of the coupled anyon plus
electromagnetic field system and Hall effects}\label{Bacry}

Having established our enlarged planar Galilei algebra,
now we  build a new theory of anyons
interacting with (constant) external fields.
Generalizing a formula of Bacry \cite{Bacry} who argued that the
Hamiltonian should be constructed from generators of the
symmetry group,  we  consider
\begin{equation}
H'=\frac{{\vec{\cP}}^2}{2m}-\frac{e}{m}\Big(\cK_{i}E_{i}+\cJ B\Big)
    -\frac{me\theta}{2B}\vE^2,
     \label{BacryHam}
\end{equation}
where that last term is dictated by boost invariance. $H'$ is indeed
$
H'=H+\cC'.
$
  Choosing a real parameter $g$, $H'$ can be further generalized as
\begin{equation}
H'_{anom}=H+\frac{g}{2}\,\cC'=
\frac{\vP{}^2}{2m}\big(1-\frac{g}{2}e\theta B\big)-e\vE\cdot\vX
-\mu B+\frac{ge\theta}{2}\,\vP\times\vE-\frac{mge\theta}{4B}\vE{}^2
     \label{anom'Ham}
\end{equation}
where $\mu=ges_{0}/2m$.
The kinetic term gets hence a field-dependent factor;
our Hamiltonian contains, together with the usual
magnetic moment term $\mu B$ [which is here a constant],
also an anomalous term proportional to $\theta\vP\times\vE$.

  Such a theory is still symmetric w.r.t. the enlarged Galilei group
  by construction.
The equation of motion, reminiscent to Eq. (5.3) of \cite{AnAn}, is
\begin{equation}
m^*\dot{X}_{i}=\big(1-\frac{g}{2}e\theta B\big)P_{i}-
     \Big(1-\frac{g}{2}\Big)em\theta\epsilon_{ij}E_{j},
\label{enlvit}
\end{equation}
supplemented with the Lorentz force law, (\ref{Lorentz}).

When $g=2$ and  $e\theta B\neq1$,  $m\dot{X}_{i}=P_{i}$,
so that our equations describe an ordinary
charged particle in an electromagnetic field.
For $g=2$ and $e\theta B=1$, Eq. (\ref{enlvit}) is identically
satisfied.

We assume henceforth that $g\neq2$.
Then Eq. (\ref{enlvit})
describes an ``exotic'' particle with anomalous moment coupling with
gyromagnetic ratio $g$, cf. \cite{AnAn}, which generalizes the
$g=0$ theory of Ref. \cite{DH}.


  Let us now consider Hall motions, i. e. such that
\begin{equation}
     \dot{X}_{i}=\epsilon_{ij}\frac{E_{j}}{B}.
     \label{Hall}
\end{equation}
For
\begin{equation}
    B\neq B''_{crit}=\frac{2}{g}\,\frac{1}{e\theta},
     \label{critB''}
\end{equation}
this is a solution of the equations of motion
(\ref{enlvit}-\ref{Lorentz}) when the momentum satisfies
\begin{equation}
     P_{i}=m\epsilon_{ij}\frac{E_{j}}{B},
     \label{PHall}
\end{equation}
The constraint (\ref{PHall}) is clearly equivalent to
{\it the vanishing of the Casimir},
$
\cC=0.
$
Conversely, from (\ref{PHall}) we infer that $\dot{P}_{i}=0$, so that
the Lorentz force on the r.h.s. of (\ref{Lorentz}) is necessarily
zero and the motion follows the Hall law.
  Thus, when $B\neq B''_{crit}$,  the Hall motions are
characterized by the constraint (\ref{PHall}), in turn equivalent
to $\cC=0$. The condition (\ref{PHall}) is invariant w. r. t.
the enlarged Galilei transformations and, when restricted to such 
motions, the
representation of the enlarged Galilei algebra becomes irreducible.

The generic motions have the familiar cycloidal form,
made of the Hall drift of the guiding center, composed
with uniform rotations with frequency
\begin{equation}
     \Omega=\frac{eB}{2m^*}\big(1-\frac{g}{2}e\theta B\big).
     \label{freq}
\end{equation}
For $g=2$ this reduces to the usual Larmor frequency $eB/m$.
The Casimir $\cC$ measures
the extent the actual motion fails to be a Hall motion.
For $m^*=0$, i. e. for
$
B=B'_{crit}=(e\theta)^{-1},
$ (\ref{enlvit}) implies
(\ref{PHall}), and hence {\it the only allowed motions are
the Hall motions} \cite{DH,AnAn}.

Interestingly, this is also what happens
for $B=B''_{crit}$ (which plainly requires $g\neq0$).
Then the momentum drops out from  (\ref{enlvit}).
   For $g=2$ (\ref{enlvit}) holds identically, but
for $g\neq2$  it becomes
\begin{equation}
    \dot{X}_{i}=\frac{g}{2}\,e\theta\epsilon_{ij}E_{j},
\label{secondary}
\end{equation}
which is once again the Hall law (\ref{Hall})
with $B=B''_{crit}$ \cite{AnAn}.
Now the constraint (\ref{PHall}) is
not enforced:
by (\ref{Lorentz}), the momentum is an arbitrary constant.

The two critical values correspond to the frequencies
$\Omega=\infty$ and $\Omega=0$, respectively. In the first case
$m^*=0$, only those initial conditions 
are consistent which satisfy (\ref{PHall}).
The system is singular and requires reduction \cite{DH}.
In the second case the initial momentum can be arbitrary,
since it has no influence on the motion. The system acquires
an extra translational symmetry in momentum space. 
\goodbreak

\section{Planar Bloch electron in external fields}\label{Bloch}

While our theory may seem to be rather speculative, it has
interesting analogies in solid state physics,
namely in the theory of a Bloch electron in a crystal.
Restricting ourselves to a single band, the
band energy and the background fields provide in fact
effective terms for the semiclassical dynamics of the
electronic wave packet \cite{Niu}.
  The mean Bloch wave vector (quasimomentum) we denote here by $\vP$
  varies in a Brillouin zone.
  In terms of $\vP$ and the mean band position coordinates, $X_{i}$,
the system is described  by the effective Lagrangian \cite{Niu}
\begin{equation}
     L^{Bloch}=P_{i}\dot{X}_{i}-\cE
     +e\big(A_{i}(\vX,t)\dot{X}_{i}+A_{0}\big)
     +\cA_{i}(\vP)\dot{P}_{i}
     \label{blochlag}
\end{equation}
with
$
A_{i}=-\half\epsilon_{ij}BX_{j},\,
A_{0}=\vE\cdot\vX.
$
The expression
$
\cE(\vP)=\cE_{0}(\vP)-M(\vP)B,
$
[where $\cE_{0}(\vP)$ is the energy of the band
and $M(\vP)$ is the mean magnetic moment]
yields  a kinetic energy term for the effective dynamics.

The (effective) vector potential $\cA_{i}$ is
the Berry connection; it can
arise, e. g., in a crystal with no spatial inversion symmetry as
in GaAs \cite{Niu}.
The last term in  (\ref{blochlag}) is indeed analogous to our 
``exotic'' term
$(\theta/2)\epsilon_{ij}P_{i}\dot{P}_{j}$ in (\ref{totlag}),
to which it would reduce  if the Berry curvature,
\begin{equation}
     \theta(\vP)=\epsilon_{ij}\frac{\ \p}{\p P_{i}}\cA_{j}(\vP),
         \label{berry}
\end{equation}
was a constant.
$\dot{X}_{i}$, physically the group velocity of
the Bloch electron, satisfies hence the equation
\begin{eqnarray}
     \Big(1-eB\theta(\vP)\Big)\dot{X}_{i}&=&\p_{P_{i}}\cE
     -e\theta(\vP)\epsilon_{ij}E_{j},
     \label{Blochvitesse}
\end{eqnarray}
supplemented with the Lorentz equation (\ref{Lorentz}).

Eq. (\ref{Blochvitesse}) has the same structure as our
original ($g=0$) velocity relation (\ref{vitesse}).
The Berry curvature
provides us with a {\it momentum dependent effective
non-commutative parameter $\theta(\vP)$}, which
yields in turn $\vP$-dependent effective mass
$
m^*=m\big(1-e\theta(P)B\big)
$
and anomalous velocity terms,
cf. \cite{Niu,AdBlo}.

In a Hamiltonian framework, the system is described by
the Bloch Hamiltonian
$H^{Bloch}=\cE(\vP)-eE_{i}X_{i}$, and by formally the same
Poisson brackets (\ref{Bcommrel}), except for the
momentum depen\-dence of $\theta$.
In particular, the mean band position coordinates do not commute,
as it had been observed a long time ago \cite{AdBlo}.
The $\vP$ dependence is consistent
with the Jacobi identity,  \cite{Berard},
even for a position-dependent $B$ \cite{DH}.

Once again, we can consider the enlarged framework of Section 
\ref{enlarged}. 
This yields the previous equations of motion and commutation 
relations,
supplemented with
$\dot{E}_{i}=0$ and $\{E_{i},\pi_{i}\}=\delta_{ij}$, respectively.
The $\vP$-dependence of NC parameter breaks the Galilean
symmetry down to the magnetic translations (times time translation) 
alone.
Assuming $\theta$ and $\cE$ only depend on
$\vP{}^2$, rotational symmetry is restored, though.
It is generated by
$\cJ$ in (\ref{enlargedangmom}), with
$\half\displaystyle\int^{\vP{}^2}\!\theta(\vP{}^2)$
replacing the ``exotic'' contribution $\theta\vP{}^2/2$.
Hence, we are  left with a residual symmetry
with generators $\cP_{i}, H,\cJ, E_{i}$ and $B$; the latter belongs
to the center. The enlarged euclidean algebra has the  Casimirs
$
\cC_{0}=BH^{Bloch}-\epsilon_{ij}\cP_{i}E_{j}
$
  and  the infinite tower $\cC_{n}=\big(\vE{}^2\big)^n,\,
n=1,\dots$

Let us now inquire about the Hall motions. Inserting the Hall law 
into 
the equation of motion (\ref{Blochvitesse}) generalizes the condition
(\ref{PHall}) as
\begin{equation}
     \p_{P_{i}}\cE=\frac{\epsilon_{ij}E_{j}}{B}.
     \label{PBlochHall}
\end{equation}

Then (\ref{PBlochHall}) satisfies $\cC=0$ for the Casimir
\begin{equation}
     \cC=B\left(\cE\big(\vP^2\big)-\cE\big(P_{0}^2\big)\right)
     -\epsilon_{ij}P_{i}E_{j}+|E|P_{0}
\label{Blochcasi}
\end{equation}
where $P_{0}$ is a solution  of the equation
$
P_{0}^2(\cE'(P_{0}^2))^2=(\vE/2B)^2.
$
That $\cC=0$ is equivalent to (\ref{PBlochHall})
 can be proved, e. g., in
the generalized parabolic case $\cE\sim(\vP{}^2)^\alpha$, 
$1/2<\alpha<3/2$. It appears, however, that the vanishing of $\cC$ is
a mere coincidence, and
 (\ref{PBlochHall}) is better interpreted as the extremum condition 
$\p_{P_{i}}\cC=0$
for the Casimir. (For (\ref{Casimir}) we obviously have a minimum).
Eq. (\ref{PBlochHall}) can have more than just one solution.
This happens, e. g. , for the energy expression 
$\cE=a^2\vP{}^2+b^2\sqrt{1+c^2\vP{}^2}$
considered by Culcer et al.  in
 Ref. \cite{AHE} for the Anomalous Hall Effect. Then
 the vanishing of $\cC$ is clearly irrelevant,
 but the extremum property is still valid.

If the effective non-commutative parameter $\theta=\theta(\vP)$
is genuinely momentum-dependent as it happens in
Ref. \cite{AHE,Niu}, the magnetic field cannot be tuned to either
of the critical values $B'_{crit}$ or $B''_{crit}$. Hence, the
Hall motions can not be made mandatory.

Writing either (\ref{vitesse}) or (\ref{Blochvitesse}) as
\begin{equation}
\dot{X}_{i}=\hbox{(kinetic term})-e\theta\epsilon_{ij}\dot{P}_{i},
\end{equation}
we recognize  the anomalous velocity  used  in the description of the
Anomalous Hall Effect in ferromagnetics \cite{AHE}.
The  additional contribution
$(g/2)em\theta\epsilon_{ij}E_{j}$ in eqn. (\ref{enlvit})  comes from
the anomalous term
  $(g/2)\theta\,\vE\times\vP$ in the Hamiltonian (\ref{anom'Ham}).
  This latter can be viewed as the ``Jackiw-Nair'' \cite{JaNa} limit
  of the spin-orbit coupling
\begin{equation}
\frac{1}{m^2c^2}\vec{\sigma}\times\vE\cdot\vP,
\label{spinorbit}
\end{equation}
advocated by Karplus and Luttinger half a century ago \cite{AHE}.
Putting $\vec{\sigma}=s\sigma_{3}$,
  $\theta=-s/c^2m^2$ (\ref{spinorbit}) becomes
indeed proportional to our anomalous term. Such a term also
arises in the non-relativistic limit of charged Dirac
particle in a constant electric field \cite{BM2}.

Adding $(ge\theta/2)\cC$ to $H^{Bloch}$ yields an anomalous extension 
of the semiclassical Bloch model with (\ref{Blochvitesse}) 
generalized to
\begin{eqnarray}
     \Big(1-eB\theta\Big)\dot{X}_{i}=\Big(1-\frac{g}{2}e\theta B\Big)
     \p_{P_{i}}\cE-\Big(1-\frac{g}{2}\Big)e\theta \epsilon_{ij}E_{j}
     -\frac{eg}{2}\cC\p_{P_{i}}\theta
     \label{Blochanomvitesse}
\end{eqnarray}
[$\theta=\theta(\vP)$],
supplemented with the Lorentz equation (\ref{Lorentz}).
For a Hall motion the momentum is constant, $\vP=\vP^0$.
Putting $\theta^{0}=\theta(\vP^0)$ and
inserting into the equation of motion 
\begin{equation}
    \Big(1-\frac{g}{2}e\theta^{0} B\Big)
    \Big(\p_{P_{i}}\cE\Big|_{\vP^0}-\epsilon_{ij}\frac{E_{j}}{B}\Big)
    \Big|_{\vP^0}=e\frac{g}{2}\cC(\vP^0)\p_{P_{i}}\theta\Big|_{\vP^0}.
    \label{newcond}
\end{equation}

Let us assume that $\p_{P_{i}}\theta\neq0$.
Although (\ref{newcond}) has more general solutions, we
observe that if the
Hall condition (\ref{PBlochHall}) holds, then we have again $\cC=0$.
At last,  $1-({g}/{2})e\theta^{0} B=0$ can also yield Hall motions of 
the second type for some particular value of the momentum.

\section{Conclusion}\label{concl}

In this Letter we presented, following  Ref. \cite{NeOl},
a framework which unifies phase space and field variables,
and used it to introduce ``enlarged Galilean symmetry''.
Then we built a theory along the lines put forward by
Bacry \cite{Bacry}. Adding a Casimir to the Hamiltonian,
our theory can accomodate  anomalous moment coupling cf. \cite{AnAn}. 
We derived an algebraic characterization
of the Hall motions and have shown that
for $B=B'_{crit}$ or $B''_{crit}$ the Hall motions become mandatory.
The physical interpretation
is provided by the semiclassical theory of
a Bloch electron, where a
momentum-dependent effective non-commutative parameter is derived
by a Berry phase calculation  \cite{Niu,Berard}.

\kikezd{Acknowledgement}.
PAH and PCS would like to thank for hospitality
the University of Lecce.  We are indebted to
 W. Zakrzewski for discussions.



\end{document}